\documentclass[journal]{IEEEtran}
\usepackage[utf8]{inputenc}
\usepackage{bm}
\usepackage{epsfig,amsfonts,amsbsy,bm,mathrsfs}
\usepackage{amssymb,amsmath,amsthm,latexsym,amscd,amsfonts}
\usepackage{lettrine}
\usepackage{authblk}
\usepackage{graphics}
\usepackage{graphicx}
\usepackage{epstopdf}
\usepackage{psfrag,float}
\usepackage{pstricks}
\usepackage{pst-plot}
\usepackage{cite}
\usepackage{balance}
\usepackage{epsfig}
\usepackage{bbm}
\usepackage{enumitem}
\usepackage{dsfont}
\usepackage{setspace}
\usepackage{float}
\usepackage{relsize}
\usepackage{comment}
\usepackage{subfigure} 
\usepackage{algorithm}
\usepackage[noend]{algpseudocode}
\usepackage[nolist]{acronym}
\usepackage{tabularx}
\usepackage{pifont}
\usepackage{units}

\usepackage{tikz}
\usetikzlibrary{calc}
\usepackage{xcolor}
\usepackage{tabularx}
\usepackage{colortbl}
\usepackage{pgfplots}
\pgfplotsset{compat=newest}
\usetikzlibrary{plotmarks}
\usetikzlibrary{arrows.meta}
\usepgfplotslibrary{patchplots}
\usepackage{grffile}
\newlength\fheight 
\newlength\fwidth 
\usepgfplotslibrary{fillbetween}

\usepackage{amsthm}

\newcommand{\herm}{^\mathrm{H}}

\newcommand{\vect}{\mathrm{vec}}

\algdef{SE}[SUBALG]{Indent}{EndIndent}{}{\algorithmicend\ }%
\algtext*{Indent}
\algtext*{EndIndent}

\makeatletter
\newcommand*\rel@kern[1]{\kern#1\dimexpr\macc@kerna}
\newcommand*\widebar[1]{%
  \begingroup
  \def\mathaccent##1##2{%
    \rel@kern{0.8}%
    \overline{\rel@kern{-0.8}\macc@nucleus\rel@kern{0.2}}%
    \rel@kern{-0.2}%
  }%
  \macc@depth\@ne
  \let\math@bgroup\@empty \let\math@egroup\macc@set@skewchar
  \mathsurround\z@ \frozen@everymath{\mathgroup\macc@group\relax}%
  \macc@set@skewchar\relax
  \let\mathaccentV\macc@nested@a
  \macc@nested@a\relax111{#1}%
  \endgroup
}
\makeatother

\newcommand{\JJ}{\boldsymbol{J}}

\newcommand{\AAb}{\boldsymbol{B}}
\newcommand{\yym}{\bm{\mathcal{Y}}}
\newcommand{\nnm}{\bm{\mathcal{N}}}
\newcommand{\dd}{\boldsymbol{d}}

\newcommand{\aab}{\boldsymbol{a}}

\newcommand{\aabx}{\aab_{\rm{x}}}
\newcommand{\aabz}{\aab_{\rm{z}}}

\newcommand{\deltaf}{\Delta_f}

\newcommand{\normsmall}[1]{\big\lVert#1\big\rVert}

\newcommand{\aabxw}{\widetilde{\aab}_{\rm{x}}}
\newcommand{\aabzw}{\widetilde{\aab}_{\rm{z}}}

\newcommand{\yymt}{\widetilde{\yym}}

\newcommand{\aabxi}{{\aab}_{{\rm{x}}}^{i}}
\newcommand{\aabzi}{{\aab}_{{\rm{z}}}^{i}}
\newcommand{\ddi}{{\dd}^{i}}

\newcommand{\Mx}{N_{\rm{x}}}
\newcommand{\Mz}{N_{\rm{z}}}
\newcommand{\dz}{d_{\rm{z}}}
\newcommand{\dx}{d_{\rm{x}}}

\newcommand{\rmel}{{{\rm{el}}}}
\newcommand{\rmaz}{{{\rm{az}}}}

\newcommand{\thetaaz}{\theta_{\rmaz}}
\newcommand{\thetael}{\theta_{\rmel}}

\makeatletter
\newcommand{\gettikzxy}[3]{%
  \tikz@scan@one@point\pgfutil@firstofone#1\relax
  \edef#2{\the\pgf@x}%
  \edef#3{\the\pgf@y}%
}
\definecolor{mygray}{gray}{0.6}

\newcommand{\complexset}[2]{ \mathbb{C}^{#1 \times #2}  }

\newcommand{\complexsett}[3]{ \mathbb{C}^{#1 \times #2 \times #3}  }

\newcommand{\abs}[1]{\big\lvert #1 \big\rvert}

\begin{acronym}[ACRONYM]
\acro{AI}{artificial intelligence}
\acro{AoA}{angle-of-arrival}
\acro{AoD}{angle-of-departure}
\acro{AWGN}{additive white Gaussian noise}
\acro{BS}{base station}
\acro{BP}{belief propagation}
\acro{CDF}{cumulative density function}
\acro{CFO}{carrier frequency offset}
\acro{CP}{cyclic prefix}
\acro{CRB}{Cram\'er-Rao bound}
\acro{CCRB}{constrained Cram\'er-Rao bound}
\acro{C-V2X}{cellular vehicle-to-anything}
\acro{DA}{data association}
\acro{D-MIMO}{distributed multiple-input multiple-output}
\acro{DL}{downlink}
\acro{DSRC}{dedicated short-range communications}
\acro{EKF}{extended Kalman filter}
\acro{KF}{Kalman filter}
\acro{EM}{electromagnetic}
\acro{FIM}{Fisher Information Matrix}
\acro{GDOP}{geometric dilution of precision}
\acro{GNSS}{global navigation satellite system}
\acro{GPS}{global positioning system}
\acro{IP}{incidence point}
\acro{IQ}{in-phase and quadrature}
\acro{ISAC}{integrated sensing and communication}
\acro{ICI}{inter-carrier interference}
\acro{JCS}{Joint Communication and Sensing}
\acro{JRC}{joint radar and communication}
\acro{JRC2LS}{joint radar communication, computation, localization, and sensing}
\acro{IMU}{inertial measurement unit}
\acro{IOO}{indoor open office}
\acro{IoT}{Internet of Things}
\acro{IRN}{infrastructure reference node}
\acro{KPI}{key performance indicator}
\acro{LoS}{line-of-sight}
\acro{LS}{least-squares}
\acro{MCRB}{misspecified Cram\'er-Rao bound}
\acro{MIMO}{multiple-input multiple-output}
\acro{SIMO}{single-input multiple-output}
\acro{ML}{maximum likelihood}
\acro{mmWave}{millimeter-wave}
\acro{NLoS}{non-line-of-sight}
\acro{NR}{new radio}
\acro{OFDM}{orthogonal frequency-division multiplexing}
\acro{OTFS}{orthogonal time-frequency-space}
\acro{OEB}{orientation error bound}
\acro{PEB}{position error bound}
\acro{VEB}{velocity error bound}
\acro{PRS}{positioning reference signal}
\acro{QoS}{Quality of Service}
\acro{RAN}{radio access network}
\acro{RAT}{radio access technology}
\acro{RCS}{radar cross section}
\acro{RedCap}{reduced capacity}
\acro{RF}{radio frequency}
\acro{RIS}{reconfigurable intelligent surface}
\acro{RFS}{random finite set}
\acro{RMSE}{root mean squared error}
\acro{MSE}{mean squared error}
\acro{RSU}{road-side unit}
\acro{RTK}{real-time kinematic}
\acro{RTT}{round-trip-time}
\acro{SLAM}{simultaneous localization and mapping}
\acro{SLAT}{simultaneous localization and tracking}
\acro{SNR}{signal-to-noise ratio}
\acro{ToA}{time-of-arrival}
\acro{TDoA}{time-difference-of-arrival}
\acro{TR}{time-reversal}
\acro{TXRX}[TX/RX]{transmitter/receiver}
\acro{Tx}{transmitter}
\acro{Rx}{receiver}
\acro{UE}{user equipment}
\acro{UL}{uplink}
\acro{ULA}{uniform linear array}
\acro{URA}{uniform rectangular array}
\acro{UWB}{ultra wideband}
\acro{V2I}{vehicle-to-infrastructure}
\acro{V2X}{vehicle-to-anything}
\acro{V2V}{vehicle-to-vehicle}
\acro{VRU}{vulnerable road user}
\acro{CRU}{connected road user}
\acro{XL-MIMO}{extra-large MIMO}
\acro{REB}{range error bound}
\acro{MIMO}{multiple-input and multiple-output}
\acro{MPC}{multipath component}
\acro{WAA}{weighted average approximation}
\acro{SVD}{singular value decomposition}
\acro{ESPRIT}{estimation of signal parameters via rotational invariance techniques}
\acro{LTE}{long-term evolution}
\acro{FR1}{frequency range 1}
\end{acronym}
\usepackage{geometry}
\begin{document}

\bibliographystyle{IEEEtran}
\bstctlcite{IEEEexample:BSTcontrol}


\title{V2X Sidelink Positioning in FR1: From Ray-Tracing and Channel Estimation to Bayesian Tracking}

\author{\IEEEauthorblockN{Yu Ge\IEEEauthorrefmark{1}, Maximilian Stark\IEEEauthorrefmark{2}, Musa Furkan Keskin\IEEEauthorrefmark{1}, Hui Chen\IEEEauthorrefmark{1},}

\IEEEauthorblockN{Guillaume Jornod\IEEEauthorrefmark{2}, Thomas Hansen\IEEEauthorrefmark{2}, Frank Hofmann\IEEEauthorrefmark{2},  Henk Wymeersch\IEEEauthorrefmark{1}}

\IEEEauthorblockA{\IEEEauthorrefmark{1}
Department of Electrical Engineering, Chalmers University of Technology, Gothenburg, Sweden, }

\IEEEauthorblockA{\IEEEauthorrefmark{2}
Robert Bosch GmbH, Hildesheim, Germany}

}

\maketitle
\pagestyle{empty}
\thispagestyle{empty}
\begin{abstract}
Sidelink positioning research predominantly focuses on the snapshot positioning problem, often within the mmWave band. Only a limited number of studies have delved into \ac{V2X} tracking within sub-6 GHz bands. In this paper, we investigate the V2X sidelink tracking challenges over sub-6 GHz frequencies. We propose a Kalman-filter-based tracking approach that leverages the estimated error covariance lower bounds (EECLBs) as measurement covariance, alongside a gating method to augment tracking performance. Through simulations employing ray-tracing data and super-resolution channel parameter estimation, we validate the feasibility of sidelink tracking using our proposed tracking filter with two novel EECLBs. Additionally, we demonstrate the efficacy of the gating method in identifying line-of-sight  paths and enhancing tracking performance.

\end{abstract}
\acresetall 
\vspace{-4mm}
\section{Introduction}
\Ac{ISAC} has garnered significant attention in the evolution of 5G mobile radio systems and emerges as a cornerstone feature in beyond 5G communication systems \cite{wild2021joint,liu2022integrated}. Many studies have  focused on the mmWave bands, primarily due to their larger available bandwidth at higher carrier frequencies, leading to enhanced time-delay resolution \cite{KanRap:21,hexax_d31}.  Lower bands, such as the sub-6 GHz bands, also hold significant promise for sensing applications,  particularly in environments with less complicated multipath propagation \cite{wild20236g}. One notable application is \ac{V2X} communication \cite{tr:38845-3gpp21}, where radio sensing supports various driving functions aimed at improving passenger safety and comfort \cite{5g_nr_v2x_driving}. Within V2X, sidelink communication can effectively utilize the 5850-5925 MHz band at \ac{FR1}, along with neighboring unlicensed bands, to bolster sensing capabilities \cite{garcia2021tutorial,lien20203gpp}, thereby enhancing integrity, accuracy, and power efficiency \cite{tr:38859-3gpp22}. As lower bands are expected to be more widely used for sidelink, it is important to understand their real-world sensing performance.


The term `sensing' in sidelink V2X encompasses not only traditional radar-like functions (e.g., mapping of landmarks, tracking of moving objects) but also  positioning of connected users \cite{hexax_d31}, which is the focus of this work. Research in positioning can be broken down into \emph{snapshot positioning} \cite{ko2021v2x,liu2021highly,fouda2021dynamic,liu2021v2x,ge2023analysis,ge2024v2x} and \emph{tracking} \cite{koivisto2017jointd,kadambi2022neural,karfakis2023nr5g,talvitie2023orientation}, where the former considers a single time step, while the latter considers 
tracking of moving users over time, taking into account their motion models. 
Several studies have explored V2X sidelink positioning within sub-6 GHz frequencies. For instance, \cite{ko2021v2x} offers a comprehensive overview of V2X positioning, highlighting the synchronization limitations of \ac{TDoA} methods and proposing the utilization of carrier phase and multipath information. Different positioning systems and architectures tailored for vehicular applications, along with relevant \acp{KPI} are described in \cite{liu2021highly}. A dynamic method for switching between \ac{GNSS} and  NR \ac{V2X} \ac{TDoA} measurements is presented in \cite{fouda2021dynamic}. Additionally, \cite{liu2021v2x} enhances positioning accuracy by combining \ac{V2V} range and angle measurements with \ac{V2I} \ac{TDoA} measurements.  In \cite{ge2023analysis,ge2024v2x}, the V2X sidelink positioning problem is investigated from an end-to-end perspective, and its theoretical performance bounds are derived.
Regarding tracking in FR1, \cite{koivisto2017jointd} demonstrates the viability of tracking by employing \ac{ToA} and \ac{AoA} estimates in dense networks. This is achieved using cascaded \acp{EKF}, where two \acp{EKF} are employed for tracking both the channel parameters of the propagation path corresponding to the largest power and the user. Notably, channel estimation is not implemented in this approach. Additionally, in \cite{kadambi2022neural}, a neural network architecture is presented, enabling joint learning of user locations over time and the surrounding environment in an unsupervised manner. This approach utilizes time of flights, obtained using the MUSIC algorithm from received signals. Furthermore, \cite{karfakis2023nr5g} conducts tracking using an EKF, which utilizes received signal strengths from many known nodes as measurements. In \cite{talvitie2023orientation}, antenna-level carrier phase measurements are harnessed within an EKF framework to continuously track users' 3D orientation and location. Synchronizing reference nodes in sidelink communication poses challenges, making solutions relying on a single reference node preferable. Surprisingly, there is limited research addressing the V2X sidelink tracking issue with a single reference node in FR1 from a standard end-to-end perspective.


In this paper, we investigate  V2X sidelink tracking in FR1, employing a \emph{comprehensive end-to-end methodology}. This approach incorporates ray-tracing-based channel modeling, a high-resolution channel estimator, and a \ac{KF}-based tracking filter. Our primary contributions are summarized as follows: (\emph{i}) We propose a novel method for estimating the measurement covariance, which is utilized in the update step of the tracking filter, by computing the estimated error covariance lower bounds (EECLBs). (\emph{ii}) We develop a gating method that utilizes tracking information to assist \ac{LoS} identification. (\emph{iii}) We validate the effectiveness of the end-to-end tracking framework using realistic ray-tracing data.


\subsubsection*{Notations} Scalars (e.g., $x$) are denoted in italic, vectors (e.g., $\boldsymbol{x}$) in bold, matrices (e.g., $\boldsymbol{X}$) in bold capital letters. Transpose is denoted by $(\cdot)^{\top}$. The Kronecker product is denoted by $\otimes$. A Gaussian density with mean $\boldsymbol{u}$ and covariance $\boldsymbol{\Sigma}$, evaluated in value $\boldsymbol{x}$ is denoted by $\mathcal{N}(\boldsymbol{x};\boldsymbol{u},\boldsymbol{\Sigma})$. A $n$-by-$n$ identical matrix is denoted by $\boldsymbol{I}_{n}$, and a $n$-by-$n$ zero matrix is denoted by $\boldsymbol{0}_{n}$.

\vspace{-2mm}
\section{System Model}
In this paper, a sub-6 GHz V2X sidelink scenario involving a single fixed \ac{RSU} and a single moving \ac{CRU} is examined. This section outlines the state models for both the RSU and the CRU, as well as the signal and measurement models.
\vspace{-4mm}
\subsection{State Models}
In the considered scenario, the fixed \ac{RSU} is equipped with a \ac{URA}. With respect to the global reference coordinate system, the center of the \ac{URA} is located at $\boldsymbol{x}_{\text{RSU}}=[x_\text{RSU},y_\text{RSU},z_\text{RSU}]^{\top}$, and its orientation is described by Euler angles $\boldsymbol{\psi}_\text{RSU}=[\varepsilon_\text{RSU},\nu_\text{RSU},\gamma_\text{RSU}]^{\top}$, ordered as roll, pitch, and yaw, respectively. Consequently, the state of the \ac{RSU} is represented as  $\boldsymbol{s}_\text{RSU}=[\boldsymbol{x}_\text{RSU}^{\top},\boldsymbol{\psi}_\text{RSU}^{\top}]^{\top}$, which remains constant over time and is assumed to be known a priori. 
The \ac{CRU} is equipped with a single antenna, positioned at $\boldsymbol{x}_{\text{CRU},k}=[x_{\text{CRU},k},y_{\text{CRU},k},z_{\text{CRU},k}]^{\top}$ at time step $k$. The \ac{CRU} moves over time, with a velocity $\dot{\boldsymbol{x}}_{\text{CRU},k}=[\dot{x}_{\text{CRU},k},\dot{y}_{\text{CRU},k},\dot{z}_{\text{CRU},k}]^{\top}$ at time step $k$ and corresponding period $T$. Consequently, the state of the \ac{CRU} at time step $k$ is encapsulated by  $\boldsymbol{s}_{\text{CRU},k}=[\boldsymbol{x}_{\text{CRU},k}^{\top},\dot{\boldsymbol{x}}^{\top}_{\text{CRU},k}]^{\top}$, which evolves according to a predetermined dynamic model given by~\cite[Ch.~13.1]{gustafsson2010statistical}
\begin{align}
    \boldsymbol{s}_{\text{CRU},k} = \boldsymbol{F}\boldsymbol{s}_{\text{CRU},k-1} + \boldsymbol{G}\boldsymbol{n}_{k},
\end{align}
where $\boldsymbol{F}\in \mathbb{R}^{6\times 6}$ represents the transition matrix, $\boldsymbol{G}\in \mathbb{R}^{6\times 2}$  and  $\boldsymbol{n}_{k}$ refers to the acceleration process noise, modeled as a zero-mean Gaussian noise with covariance $\boldsymbol{Q}=\sigma^{2}_{\text{a}}\boldsymbol{I}_{2}$, such that $\boldsymbol{n}_{k} \sim \mathcal{N}(\boldsymbol{n}_{k};\boldsymbol{0},\boldsymbol{Q})$. Therefore, the transition density can be characterized as
\begin{equation}
p(\boldsymbol{s}_{\text{CRU},k} | \boldsymbol{s}_{\text{CRU},k-1}) = {\cal N}(\boldsymbol{s}_{\text{CRU},k} ; \boldsymbol{F}\boldsymbol{s}_{\text{CRU},k-1},\boldsymbol{G}\boldsymbol{Q}\boldsymbol{G}^{\top}). \label{int_dynamicmodel}
\end{equation}
\vspace{-4mm}
\subsection{Signal Model}
We follow a \ac{RTT} protocol in this paper. Every time step, the \ac{RSU} sends \ac{OFDM} downlink pilot signals to the \ac{CRU}, comprising $N_{\text{OFDM}}$ symbols over $S$ subcarriers. These signals can reach the \ac{CRU} via the \ac{LoS} path, which is the path that signals reach \ac{CRU} directly and \ac{NLoS} paths which are the paths that signals bounce off landmarks in the environment and then reach the \ac{CRU}. The \ac{CRU} responds with signals to the \ac{RSU} following the \ac{RTT} protocol. Additionally, we make the assumption that the transmission interval is sufficiently short to neglect any Doppler-induced phase effects. Consequently, the received signal at time step $k$ for the $g$-th \ac{OFDM} symbol at the $\kappa$-th subcarrier can be expressed as \cite{heath2016overview}
\begin{align}
    \boldsymbol{y}_{\kappa,g,k}&=  {{\sum _{i=0}^{I_{k}-1}\rho^{i}_{k}\boldsymbol{a}(\boldsymbol{\theta}_{k}^{i})e^{-\jmath 2\pi \kappa \Delta_f \tau_{k}^{i}}}}  x_{\kappa,g} +  \boldsymbol{\omega}_{\kappa,g,k},\label{eq:signalModel} 
\end{align}
where $x_{\kappa,g}$ is the pilot signal, $\boldsymbol{y}_{\kappa,g,k}$ is the received signal across the \ac{RSU} array,  $\boldsymbol{\omega}_{\kappa,g,k}$ is the \ac{AWGN}, $\Delta_f$ is the subcarrier spacing, and $\boldsymbol{a}(\cdot)$ is the steering vectors of the \ac{RSU} antenna arrays. There are $I_{k}$ paths from landmarks in the environment at time step $k$ with $i=0$ representing the \ac{LoS} path. Each path $i$ can be described by a complex gain $\rho_{k}^{i}$, a \ac{ToA} $\tau_{k}^{i}$, and an \ac{AoA} pair $\boldsymbol{\theta}_{k}^{i}=[{\theta}_{\text{az},k}^{i},{\theta}_{\text{el},k}^{i}]^{\top}$ in azimuth and elevation. Please note that the signal model \eqref{eq:signalModel} is valid for both the \ac{RSU} and \ac{CRU} sides when acting as a receiver. The \ac{AoA} pair $\boldsymbol{\theta}_{k}^{i}$ is determined by the direction of the signals that arrive the receiver (\ac{RSU} or \ac{CRU}) for the $i$-th path, and the \ac{ToA} $\tau_{k}^{i}$ is determined by the propagation distance of the $i$-th path and a clock offset between the transmitter and the receiver (cancelled out through the RTT protocol). 

\begin{figure}
    \centering
    \includegraphics[width=1\columnwidth]{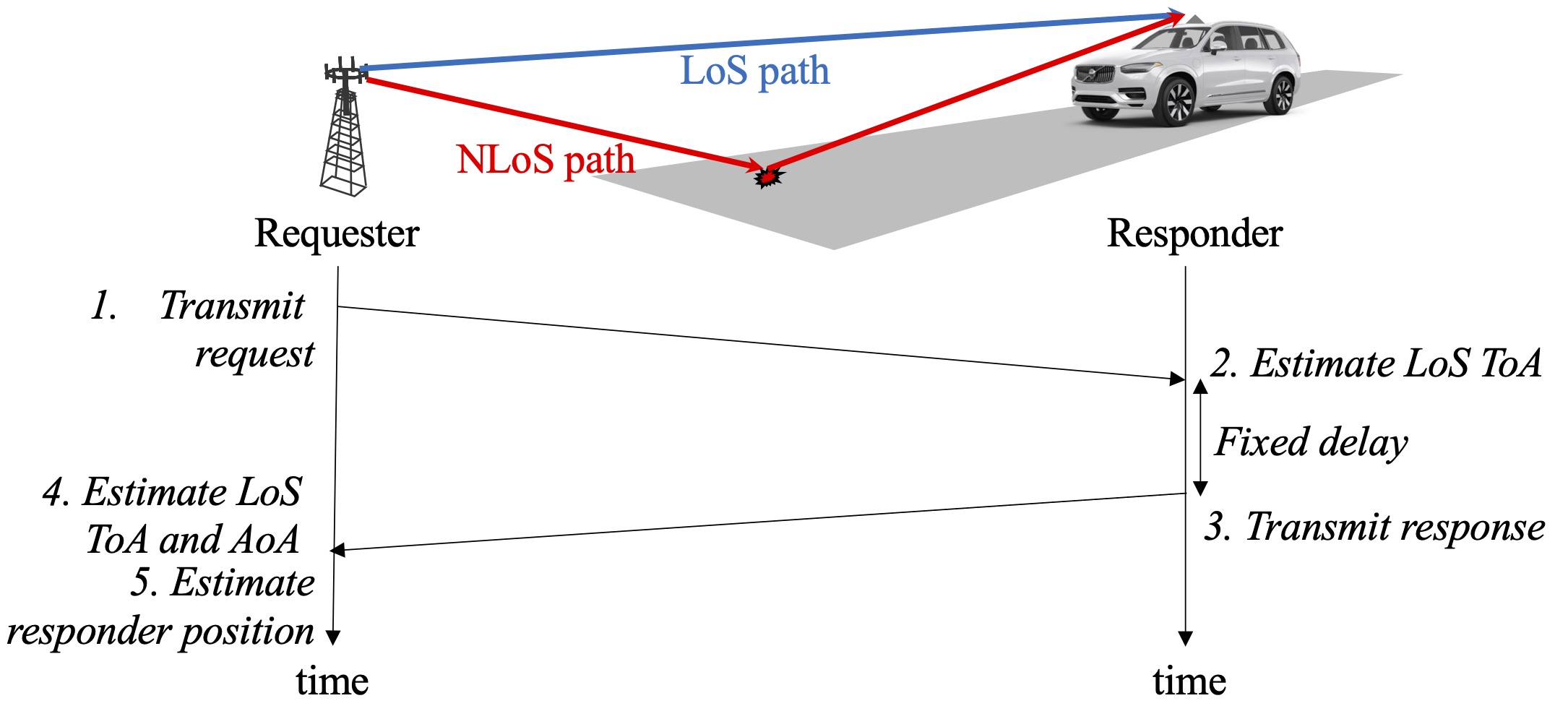}
    \caption{The considered \ac{RTT} protocol between a requester (a \ac{RSU}) and a responder (a \ac{CRU}). }\vspace{-5mm}
    \label{fig:RTTprotocol}
\end{figure}

The \ac{LoS} \ac{AoA} $\boldsymbol{\theta}_{k}^{0}$ at the \ac{RSU} can be directly related to $\boldsymbol{x}_{\text{CRU},k}$ (see \cite{ge2024v2x}), but the \ac{LoS}  \ac{ToA} is given by  $\tau_{k}^{0}=\Vert \boldsymbol{x}_{\text{RSU}} -\boldsymbol{x}_{\text{CRU},k}\Vert/c +b_{\text{CRU},k}$, which depends on an unknown time-varying clock bias $b_{\text{CRU},k}$ between the \ac{RSU} and \ac{CRU}. Here $c$ denotes the speed of light. To avoid tracking this bias, 
we utilize the \ac{RTT} protocol, which is often used in range-based positioning to deal with the clock offset \cite{5g_nr_rel16,5g_nr_v2x_driving}. As shown in Fig.~\ref{fig:RTTprotocol}, 
the \ac{RSU} can measure the \ac{RTT} to the \ac{CRU} and infer the distance by subtracting the known processing time at the \ac{CRU}.

\subsection{Measurement Model}
A channel estimator (see Section \ref{channel_estimator}) will be applied on \eqref{eq:signalModel} to recover the channel parameters, i.e., gain, \ac{ToA} and \ac{AoA}. Channel parameters provided by the channel estimator at time step $k$ are denoted by
$\mathcal{G}_{k}=\{\boldsymbol{g}_{k}^{0},\dots, \boldsymbol{g}_{k}^{\hat{{I}}_{k}-1}\}, \forall l \in \{0,\dots,\hat{{I}}_{k}-1\}$, with $\boldsymbol{g}_{k}^{l}=[\hat{\rho}_{k}^{l},\hat{\tau}_{k}^{l},(\hat{\boldsymbol{\theta}}_{k}^{l})^{\top}]^{\top}$ denoting an estimated path. In general, $\hat{{I}}_{k}\neq {I}_{k}$, since (i) not all paths can be resolved and some paths may be missed (including the \ac{LoS} path); (ii) extra entries could be introduced in $\mathcal{G}_{k}$ due to noise peaks during channel estimation. 
Hence, neither the existence of the \ac{LoS} in $\mathcal{G}_{k}$ nor the index of that path is known.

To  track the CRU using the \ac{LoS} path, we determine the \ac{LoS} path in $\mathcal{G}_{k}$ and convert it to a position estimate with an associated measurement covariance. 
Hence, we directly use the estimated position as measurement, which is denoted as $\boldsymbol{z}_{k}$ at time $k$. Consequently, we have the measurement function as
\begin{align}
    \boldsymbol{z}_{k}= \boldsymbol{H}\boldsymbol{s}_{\text{CRU},k} + \boldsymbol{r}_{k},
\end{align}
where $\boldsymbol{H}$ is the observation matrix describing the measurement function from $\boldsymbol{s}_{\text{CRU},k}$ to $\boldsymbol{z}_{k}$. Since $\boldsymbol{z}_{k}$ is the measurement of \ac{CRU} position $\boldsymbol{x}_{\text{CRU},k}$ only, $\boldsymbol{H}=[\boldsymbol{I}_{3},\boldsymbol{0}_{3}]$. Moreover, $\boldsymbol{r}_{k}$ refers to the measurement noise, modeled as a zero-mean Gaussian noise with covariance $\boldsymbol{R}_{k}$, such that $\boldsymbol{r}_{k} \sim \mathcal{N}(\boldsymbol{r}_{k};\boldsymbol{0},\boldsymbol{R}_{k})$. Therefore, the likelihood can be described as
\begin{equation}
p(\boldsymbol{z}_{k} | \boldsymbol{s}_{\text{CRU},k}) = {\cal N}(\boldsymbol{z}_{k} ; \boldsymbol{H}\boldsymbol{s}_{\text{CRU},k},\boldsymbol{R}_{k}). \label{likelihood}
\end{equation}

\vspace{-3mm}

\section{End-to-end Methodology}
To solve the CRU tracking problem, the formulation provided in \eqref{int_dynamicmodel} and \eqref{likelihood} aligns well with a \ac{KF} approach. In this section, we will detail the \ac{KF} and its challenges in this application.  To address these challenges, we utilize the EECLBs (introduced in Section \ref{Sec:bounds}) as the measurement covariance $\boldsymbol{R}_{k}$, and the predicted density ${\cal N}(\boldsymbol{s}_{\text{CRU},k} ; \boldsymbol{m}_{k|k-1},\boldsymbol{P}_{k|k-1})$ is used to improve the \ac{LoS} identification (introduced in Section \ref{Sec:los_identification}).

\vspace{-3mm}

\subsection{Tracking Filter}\label{sec_track}
The problem of estimating $\boldsymbol{s}_{\text{CRU},k}$ over time through noisy measurements $\boldsymbol{z}_{1:k}$ can be posed as a Bayesian filtering problem. 
The \ac{KF} offers a closed-form solution to the Bayesian filtering equations when dealing with linear Gaussian dynamic and measurement models \cite[Ch. 7.1]{gustafsson2010statistical}. As all predicted and posterior densities are Gaussian, we denote the predicted and posterior densities at time step $k$ as
\begin{align}
f(\boldsymbol{s}_{\text{CRU},k} | \boldsymbol{s}_{\text{CRU},k-1},\boldsymbol{z}_{1:k-1}) &= {\cal N}(\boldsymbol{s}_{\text{CRU},k} ; \boldsymbol{m}_{k|k-1},\boldsymbol{P}_{k|k-1}), \nonumber\\
f(\boldsymbol{s}_{\text{CRU},k}|\boldsymbol{z}_{1:k}) &= {\cal N}(\boldsymbol{s}_{\text{CRU},k} ; \boldsymbol{m}_{k|k},\boldsymbol{P}_{k|k}), \nonumber
\end{align}
with $\boldsymbol{m}_{k|k-1}$ and $\boldsymbol{P}_{k|k-1}$ denoting the mean and the covariance of $f(\boldsymbol{s}_{\text{CRU},k} | \boldsymbol{s}_{\text{CRU},k-1},\boldsymbol{z}_{1:k-1})$, and  $\boldsymbol{m}_{k|k}$, and $\boldsymbol{P}_{k|k}$ denoting the mean and the covariance of $f(\boldsymbol{s}_{\text{CRU},k}|\boldsymbol{z}_{1:k})$. The predicted and posterior density parameters are \cite[Ch. 7.1]{gustafsson2010statistical}:
\begin{itemize}
    \item \emph{Prediction step:} The predicted mean and covariance are 
    \begin{align} \label{eq_pred_mean}
&\boldsymbol{m}_{k|k-1}=\boldsymbol{F}\boldsymbol{m}_{k-1|k-1},\\ \label{eq_pred_cov}
&\boldsymbol{P}_{k|k-1}=\boldsymbol{F}\boldsymbol{P}_{k-1|k-1}\boldsymbol{F}^{\top} + \boldsymbol{G}\boldsymbol{Q}\boldsymbol{G}^{\top}.
\end{align}
     \item \emph{Update step:} From the measurement $\boldsymbol{z}_{k}$, the updated mean and covariance are obtained as
     \begin{align}
&\hat{\boldsymbol{z}}_{k}=\boldsymbol{H}\boldsymbol{m}_{k|k-1}\label{Kamlam_update1},\\
&\boldsymbol{S}_{k}=\boldsymbol{H}\boldsymbol{P}_{k|k-1}\boldsymbol{H}^{\top} + \boldsymbol{R}_{k},\\
&\boldsymbol{K}_{k}=\boldsymbol{P}_{k|k-1}\boldsymbol{H}^{\top}\boldsymbol{S}_{k}^{-1},\\
&\boldsymbol{m}_{k|k}=\boldsymbol{m}_{k|k-1}+\boldsymbol{K}_{k}(\boldsymbol{z}_{k}-\hat{\boldsymbol{z}}_{k}),\\
&\boldsymbol{P}_{k|k}=\boldsymbol{P}_{k|k-1}-\boldsymbol{K}_{k}\boldsymbol{H}\boldsymbol{P}_{k|k-1},\label{Kamlam_update5}
\end{align}
where $(\boldsymbol{z}_{k}-\hat{\boldsymbol{z}}_{k})$, $\boldsymbol{S}_{k}$ and $\boldsymbol{K}_{k}$ are usually referred as the innovation, innovation covariance, and Kalman gain, respectively. Note that the update step is skipped if no measurement arrives, i.e., the channel estimator does not provide an estimation of the \ac{LoS} path.
\end{itemize}
The recursion begins with the mean $\boldsymbol{m}_{0|0}$ and the covariance $\boldsymbol{P}_{0|0}$ of the prior density $f(\boldsymbol{s}_{\text{CRU},0})$. 

The main challenges of implementing the tracking filter under the considered FR1 scenario are as follows:
\begin{itemize}
 \item \emph{Measurement covariance estimation:} The measurement covariance $\boldsymbol{R}_{k}$ reflects the quality of the measurement and affects how much the tracking filter should believe in the measurement. However, $\boldsymbol{R}_{k}$ is not directly available in the considered problem as the receiver only has access to \eqref{eq:signalModel}. Therefore, a good estimation on $\boldsymbol{R}_{k}$ is essential to have good tracking performance.

\item \emph{\ac{LoS} identification:} Channel estimator provides a group of estimated paths every time step.
We do not know  if the \ac{LoS} path is part of $\mathcal{G}_{k}$, nor what its index is. 
Hence, we should identify which is the estimation of the \ac{LoS} path and use the position estimate from it as the measurement. The estimated path with the shortest delay is usually chosen as the \ac{LoS} path. However, this method can lead to large errors since the shortest estimated path does not always correspond to the \ac{LoS} path.
\end{itemize}


\vspace{-3mm}

\subsection{Measurement Noise Covariance Estimation} \label{Sec:bounds}
In this section, the fundamentals of \ac{FIM} are briefly introduced, and two proposed EECLB to estimate $\boldsymbol{R}_{k}$ are described.

\subsubsection{\ac{FIM}}

\ac{FIM} offers a prevalent method for assessing the performance of estimation algorithms in statistical estimation theory, indicating the \ac{MSE} of any unbiased estimator should always be larger than the inverse of the \ac{FIM} \cite[Ch. 4.2]{van2004detection}. Suppose we have a measurement model $ \boldsymbol{\psi}=\boldsymbol{\mu}(\boldsymbol{\eta},\boldsymbol{\kappa}) + \boldsymbol{\varpi}$, with $\boldsymbol{\varpi}$ denoting a complex white noise with variance $N_0$, $\boldsymbol{\mu}(\cdot)$ denoting a known possibly non-linear mapping, $\boldsymbol{\eta}$ denoting the parameters of interest, and $\boldsymbol{\kappa}$ denoting the nuisance parameters. The lower bound on the estimation error of $\boldsymbol{\xi}=[\boldsymbol{\eta}^{\top},\boldsymbol{\kappa}^{\top}]^{\top}$ has the form
\begin{align}
    \mathbb{E} \big \{ (\boldsymbol{\xi}-\hat{\boldsymbol{\xi}})((\boldsymbol{\xi}-\hat{\boldsymbol{\xi}}))^\top \big\} \succeq  \boldsymbol{\Xi}^{-1}(\boldsymbol{\xi}), \label{eq:FIMCRB}
\end{align}
where $\hat{\boldsymbol{\xi}}$ denotes  an estimate of $\boldsymbol{\xi}$, and $\boldsymbol{\Xi}(\boldsymbol{\xi})$ is the \ac{FIM} of $\boldsymbol{\xi}$ given by
\begin{align}
    \boldsymbol{\Xi}(\boldsymbol{\xi})=\frac{2}{N_0} \Re\Big \{ \big(\frac{\partial \boldsymbol{\mu}(\boldsymbol{\xi})}{\partial \boldsymbol{\xi}}\big)^{\mathsf{H}}
    \frac{\partial \boldsymbol{\mu}(\boldsymbol{\xi})}{\partial \boldsymbol{\xi}}\Big\}, \label{FIM_compute}
\end{align}
with ${\partial \boldsymbol{\mu}(\boldsymbol{\xi})}/{\partial \boldsymbol{\xi}}$ denoting the gradient matrix. The inequality in \eqref{eq:FIMCRB} indicates that the \ac{MSE} of an estimator is larger than $\boldsymbol{\Xi}^{-1}$ in the positive semidefinite sense. The  error covariance lower bound (ECLB) of $\boldsymbol{\eta}$ can be obtained by taking the corresponding sub-matrix of $\boldsymbol{\Xi}^{-1}(\boldsymbol{\xi})$, i.e., $\text{ECLB}(\boldsymbol{\eta})=[\boldsymbol{\Xi}^{-1}(\boldsymbol{\xi})]_{1:\text{dim}(\boldsymbol{\eta}),1:\text{dim}(\boldsymbol{\eta})}$, where $\text{dim}(\cdot)$ returns the dimension of the argument.

\subsubsection{{Measurement  Covariance Estimation}}
{We use estimates of the ECLB of the \ac{CRU} position to estimate the measurement noise covariance.} As only the \ac{LoS} paths are used in this problem, we compute the ECLB in two steps. Firstly, we compute the ECLB of $\boldsymbol{\eta}=[\tau^0_{k},(\boldsymbol{\theta}^0_{k})^{\top}]^{\top}$ related to the observation \eqref{eq:signalModel} according to \eqref{FIM_compute} every time step. Secondly, the ECLB of the \ac{CRU} position can be computed using a standard Jacobian approach, as the $[\tau^0_{k},(\boldsymbol{\theta}^0_{k})^{\top}]^{\top}$ is a function of $\boldsymbol{x}_{\text{CRU},k}$. However,  computing the ECLB  requires knowing the ground-truth channel parameters and the ground-truth \ac{CRU} position, which are not available in real implementation. To estimate the ECLB, we view the estimated paths $\mathcal{G}_{k}$ provided by the channel estimator and the predicted mean of the \ac{CRU} state $\boldsymbol{m}_{k|k-1}$ as ground-truth, and use these values into ECLB computation. Similar to the \ac{LoS} and the \ac{NLoS} bounds in \cite{ge2024v2x}, we compute the estimations of two ECLBs:
\begin{itemize}
    \item \emph{ECLB based on the estimated \ac{LoS} parameters}: only the \ac{LoS}  path in $\mathcal{G}_{k}$ is utilized in \eqref{eq:signalModel}, in both links of the \ac{RTT} protocol. The resulting estimated ECLB (EECLB) is denoted by EECLB-LOS.  
    \item \emph{ECLB based on the estimated \ac{LoS} and \ac{NLoS} parameters}: all paths in $\mathcal{G}_{k}$ are utilized in \eqref{eq:signalModel}, in both links of the \ac{RTT} protocol. Given that paths outside the resolution cell of the \ac{LoS} path do not impact the estimation accuracy of the \ac{LoS} path, we omit these paths when computing the ECLB. The resulting EECLB is denoted by EECLB-NLOS.  
\end{itemize}
These EECLBs are directly used as $\boldsymbol{R}_{k}$ in the tracking filter. Please note if there is no estimate within $\mathcal{G}_{k}$ corresponding to the \ac{LoS} path, then neither of the two EECLBs exists.   
\vspace{-3mm}
\subsection{\ac{LoS} Identification}

To understand the \ac{LoS} identification, we first describe the channel parameter estimator, which generates $\mathcal{G}_{k}$ from \eqref{eq:signalModel}. 

\subsubsection{Channel Parameter Estimation}


We describe the high-resolution channel estimation algorithm used to infer the geometric path parameters (i.e., gain, delay and AoA) using the observations in \eqref{eq:signalModel}. 
In \eqref{eq:signalModel}, we wipe off the constant modulus pilots $x_{\kappa,g}$ via conjugate multiplication and coherently integrate over $N_{\text{OFDM}}$ symbols, assuming short transmission interval (i.e., Doppler-induced phases are ignored). Then, \eqref{eq:signalModel} can be reformulated as a 3-D tensor $\yym_k \in \complexsett{\Mz}{\Mx}{S}$ 
\begin{align} \label{eq_yy_3d}
     \yym_k = \sum_{i=0}^{I_k-1} \rho^{i}_{k} \aabz(\boldsymbol{\theta}_{k}^{i}) \circ \aabx(\boldsymbol{\theta}_{k}^{i}) \circ \dd(\tau_{k}^{i}) + \nnm_k \,, 
\end{align}
where $\circ$ denotes the outer product and integration gains are absorbed in $\rho^{i}_{k}$. In \eqref{eq_yy_3d}, the horizontal and vertical spatial-domain steering vectors, $\aabx(\boldsymbol{\theta}) \in \complexset{\Mx}{1}$ and $\aabz(\boldsymbol{\theta}) \in \complexset{\Mz}{1}$, are defined, respectively, as
 \begin{align} \nonumber 
    [\aabx(\boldsymbol{\theta})]_n &= e^{\jmath \frac{2 \pi}{\lambda}   \dx n \cos(\thetael) \sin(\thetaaz)} ~,
    [\aabz(\boldsymbol{\theta})]_n 
    = e^{\jmath \frac{2 \pi}{\lambda}   \dz n \sin(\thetael)} ~,
\end{align}
where $\dx$ and $\dz$ denote the element spacing, $\Mx$ and $\Mz$ are the number of antenna elements in the horizontal and vertical axes of the \ac{URA}, 
and $\aab(\boldsymbol{\theta}) = \aabx(\boldsymbol{\theta}) \otimes \aabz(\boldsymbol{\theta})$. In addition, $\dd(\tau) \in \complexset{S}{1}$ represents the frequency-domain steering vector with $[\dd(\tau)]_{\kappa}=e^{-\jmath2\pi \kappa\Delta_{f}\tau}$ and $\nnm_k$ is the noise tensor.

To estimate the path parameters from \eqref{eq_yy_3d}, we employ a tensor decomposition based high-resolution channel estimation algorithm \cite{ge2024v2x}. We first construct a new tensor $\yymt_k \in \complexsett{\Mz(n_z+1)}{\Mx(n_x+1)}{V} $ (where $V = S-n_z-n_x$) from the original tensor observation in \eqref{eq_yy_3d} by augmenting the spatial domain with the frequency domain measurements \cite[Eq.~(15)]{ge2024v2x} to deal with the rank deficiency problem in standard CP decomposition (CPD)-based channel estimation. After applying this spatial augmentation (SA) strategy, we have  
\begin{align} \label{eq_yymt_3d}
     &\widetilde{\yym}_k = \sum_{i=0}^{I_k-1}\rho^{i}_{k} \aabzw(\boldsymbol{\theta}_{k}^{i},\tau_k^{i}) \circ \aabxw(\boldsymbol{\theta}_{k}^{i},\tau_k^{i}) \circ \widetilde{\dd}(\tau_k^{i}) + \widetilde{\nnm}_k
 \end{align}
 where
 \begin{subequations} \label{eq_sa_steering}
 \begin{align}
     \aabzw(\boldsymbol{\theta},\tau) &= \aabz(\boldsymbol{\theta}) \otimes  [\dd(\tau)]_{1:n_z+1} \in \complexset{\Mz(n_z+1)}{1} ~, \\
     \aabxw(\boldsymbol{\theta},\tau) &= \aabx(\boldsymbol{\theta}) \otimes  [\dd(\tau)]_{1:n_z+1} \in \complexset{\Mx(n_x+1)}{1} ~, \\
     \widetilde{\dd}(\tau) &= [\dd(\tau)]_{1:V}  \in \complexset{V}{1}~.
 \end{align}
 \end{subequations} 

Next, 
\label{channel_estimator}
we apply CPD to $\widetilde{\yym}_k$ in \eqref{eq_yymt_3d}, i.e.,
\begin{subequations} \label{eq_cp_sa}
\begin{align}
    \min_{ \{\aabzi, \aabxi, \ddi \}_{i=0}^{I_k-1} } & ~\normsmall{\hat{\yym} - \widetilde{\yym}_k}_F^2
    \\
    \mathrm{s.t.}&~~  \hat{\yym} = \sum_{i=0}^{I_k-1} \aabzi \circ \aabxi \circ \ddi ~.
\end{align}
\end{subequations}  
Based on \eqref{eq_sa_steering}, the parameters of the $I_k$ paths at time step $k$ can be estimated from the output of \eqref{eq_cp_sa} using
{\allowdisplaybreaks
\begin{subequations}\label{eq_cpd_sa_est}
\begin{align} \label{eq_tauhatl_cpd_sa}
 \hat{\tau}_k^{i} &=  -\frac{
    \angle \big( [\JJ_{1} \ddi]\herm [\JJ_{2} \ddi]  \big)    }{ 2 \pi \deltaf} ~, \\
    \hat{\theta}^{i}_{\text{el},k}&= \arg \max_{\thetael} \abs{ (\aabzi)\herm  \aabzw(\boldsymbol{\theta}, \hat{\tau}_k^{i})   }^2 ~,    \\
    \hat{\theta}^{i}_{\text{az},k} &= \arg \max_{\thetaaz} \abs{ (\aabxi)\herm  \aabxw(\thetaaz, \hat{\theta}^{i}_{\text{el},k}, \hat{\tau}_k^{i})   }^2 ~,
\end{align}
\end{subequations}}and $\hat{\boldsymbol{\rho}}_{k} = \AAb_k^{\dagger} \breve{\boldsymbol{y}}_k$,  where $\JJ_{1}$ and $\JJ_{2}$ are selection matrices selecting the first and the last $V-1$ elements of a given right-multiplying vector, $\AAb_k \in \complexset{\Mz \Mx S}{I_k}$ whose $i$-th column is given by $[\AAb_k]_{:,i} = \dd(\hat{\tau}_k^{i}) \otimes \aabx(\hat{\theta}^{i}_{\text{az},k}, \hat{\theta}^{i}_{\text{el},k}) \otimes \aabz(\hat{\theta}^{i}_{\text{az},k},\hat{\theta}^{i}_{\text{el},k}) $, $\breve{\boldsymbol{y}}_k = \vect\left(\yym_k\right) \in \complexset{\Mz \Mx S}{1}$, $\hat{\boldsymbol{\rho}}_{k} \in \complexset{I_k}{1} $ with $[\hat{\boldsymbol{\rho}}_{k}]_i = \hat{\rho}_{k}^{i}$ denoting the estimate of the path gain $\rho^{i}_{k}$ in \eqref{eq_yy_3d}, and $(\cdot)^{\dagger}$ denotes Moore-Penrose pseudo-inverse.

\subsubsection{Tracking-aided \ac{LoS} Identification} \label{Sec:los_identification}
At time step $k$, we view the predicted \ac{CRU} position in \eqref{eq_pred_mean} as the ground-truth position. From this, we reconstruct ${\tau}_{k}^{0}$ and ${\boldsymbol{\theta}}_{k}^{0}$, denoted as $\check{\boldsymbol{q}}_{k}=[\check{\tau}_{k}^{0},(\check{\boldsymbol{\theta}}_{k}^{0})^{\top}]^{\top}$. Subsequently, we define a range-angle gate centered around $\check{\boldsymbol{q}}_{k}$, with the spread determined by $\boldsymbol{P}_{k|k-1}$, i.e.,
\begin{align}\label{eq_gate}
    \mathcal{H}_k = \{ \boldsymbol{q} ~ | ~ (\boldsymbol{q}-\check{\boldsymbol{q}}_{k})^{\top}(\boldsymbol{U}_{k})^{-1}(\boldsymbol{q}-\check{\boldsymbol{q}}_{k}) \leq \beta \},
\end{align}
where $\beta$ is the gating threshold, $\boldsymbol{U}_{k}=\boldsymbol{M}_{k}\boldsymbol{P}_{k|k-1}\boldsymbol{M}_{k}^{\top}$ approximates the covariance of $\check{\boldsymbol{q}}_{k}$ with $\boldsymbol{M}_{k}$ denoting the Jacobian from the predicted position to $\check{\boldsymbol{q}}_{k}$. Using the gate defined in \eqref{eq_gate}, the detection of \ac{LoS} occurs if at least one of the estimated paths falls within the gate $\mathcal{H}_k$. Then, among those paths, the one with the closest \ac{ToA} to $\check{\tau}_{k}^{0}$ is selected as the \ac{LoS} path. If no \ac{LoS} path is found, the update step in the \ac{KF} is skipped.

\vspace{-4mm}
\section{Simulations} \label{simulation}

\subsection{Simulation Environment}
\begin{figure}
    \centering
    \includegraphics[width=0.75\linewidth]{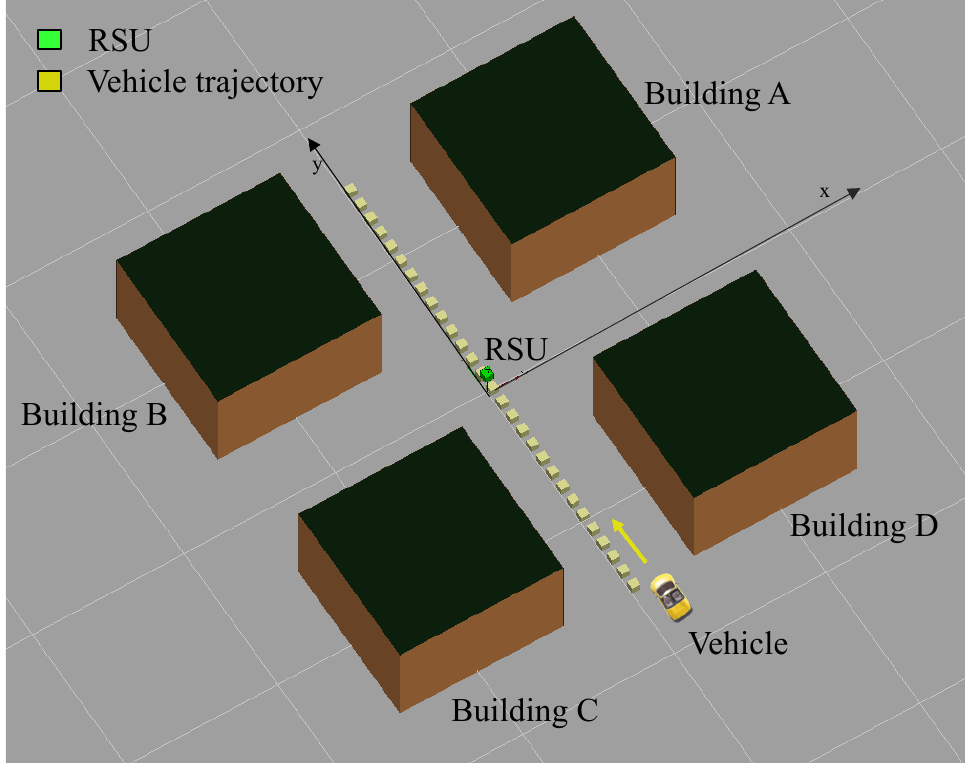}
    \vspace{-2mm}
    \caption{The ray-tracing simulation environment with a single \ac{RSU} at the center of the intersection,  four buildings, and a vehicle moving from down to up.}
    \label{fig:raytracing_env}\vspace{-6mm}
\end{figure}

We adhere to the 3GPP \ac{RSU} deployment procedure according to \cite{tr:37885-3gpp19}. We focus on an urban vehicular scenario situated at an intersection, depicted in Fig.~\ref{fig:raytracing_env}, and simulate it using the REMCOM Wireless InSite\textregistered  ray-tracer \cite{WirelessInSite}. The \ac{RSU} is located at the center of the intersection, precisely at coordinates $[0 \, \text{m},0 \, \text{m},10\, \text{m}]^{\top}$. Surrounding the intersection are four buildings, with their centers located at $[\pm 45 \, \text{m},\pm 45 \, \text{m},15\, \text{m}]^{\top}$, respectively, and all measuring $50 \, \text{m}$ in length, $50 \, \text{m}$ in width, and $30 \, \text{m}$ in height. The ground material is concrete, while the buildings feature brick walls. The \ac{RSU} is equipment with a $2\times4$ antenna array, and the \ac{CRU} is deployed with a single antenna, following the antenna patterns specified in \cite{ge2024v2x}. The \ac{CRU} possesses an initial velocity along the y-axis, and travels on the ground alongside the lane,  with its antenna positioned at a height of $1.5\, \text{m}$. The center of lane extends from $[1.6 \, \text{m},-70 \, \text{m},0 \, \text{m}]^{\top}$  to $[1.6 \, \text{m},70 \, \text{m},0 \, \text{m}]^{\top}$.  

In terms of signal parameters, the carrier frequency is set at $5.9~\text{GHz}$. The OFDM pilot signals consist of $12$ symbols, each maintaining a constant amplitude. The total active bandwidth spans $17.28~\text{MHz}$, distributed over $288$ subcarriers with a subcarrier spacing of $60~\text{kHz}$. The transmit power is configured to $10~\text{dBm}$, the noise spectral density stands at $-174~\text{dBm/Hz}$, and the receiver noise figure is set at $8~\text{dB}$. Measurements are generated every 100 ms. Additionally, we assume a 2D constant velocity model for \eqref{int_dynamicmodel}, with $\sigma_{\text{a}}=0.1~\text{m}/\text{s}^{2}$ and  $T=10~\text{ms}$. Despite the scenario being 3D, we operate in a 2D context, as the CRU traverses the ground and maintains its antenna height fixed at $1.5\, \text{m}$.

We assess the tracking performance by computing the \ac{RMSE} of the \ac{CRU} position estimates across 100 Monte Carlo (MC) simulations or \acp{CDF} of absolute errors. In each simulation, we first generate propagation paths using the raytracing data, then employ the channel estimator introduced in Section \ref{channel_estimator} on the generated signals to acquire channel parameters. Subsequently, we derive a position estimate from the identified \ac{LoS} path and apply the tracking filter 
to track the \ac{CRU} over time. We compare RMSE performances using EECLB-LOS and EECLB-NLOS as measurement covariance against four benchmarks (BMs): {\textbf{BM1: tracking using non-stationary \ac{MSE}} as the measurement covariance (namely tracking, non-stat.), which is the \textit{instantaneous} position \ac{MSE} estimated across 100 MC simulations for each time step/location; \textbf{BM2: tracking using stationary \ac{MSE}} as the measurement covariance (namely tracking, stat.), which is the \textit{average} position \ac{MSE}, i.e., averaged across 100 MC simulations and across all time steps/locations; \textbf{BM3: snapshot positioning} (position estimates from the identified \ac{LoS} paths \textit{without any tracking}), and \textbf{BM4: dead reckoning} (pure integration \textit{without considering measurements}). Later, we will also  compare tracking performances of EECLB-LOS and EECLB-NLOS in cases with and without gating}.

\vspace{-4mm}

\subsection{Results and Discussion}
\begin{figure}
\center
\input{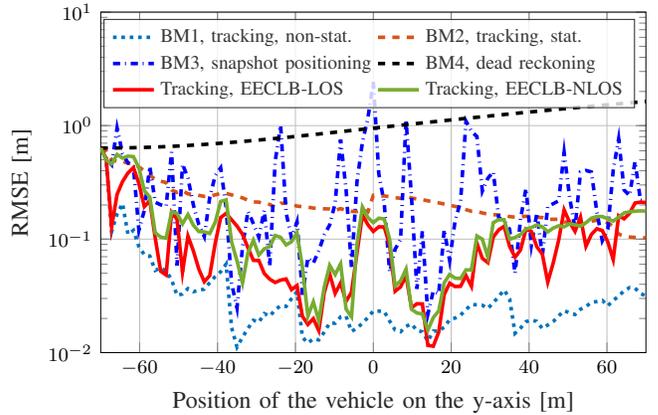}
\vspace{-3mm}
\caption{Comparison of tracking performances when EECLB-LOS and EECLB-NLOS are used as the measurement covariance in the tracking filter with four benchmarks. Here, gating is not implemented in \ac{LoS} identification.}
\label{Withoutgating}
\vspace{-4mm} 
\end{figure}

We begin by examining the viability of the tracking filter employing the proposed covariance estimates. Fig.~\ref{Withoutgating} shows the tracking results using EECLB-LOS and EECLB-NLOS as the measurement covariance, compared against four benchmarks, where the gating is not implemented. Our observations indicate that the tracking filter performs admirably, generally showcasing lower \acp{RMSE} compared to the snapshot positioning and dead reckoning results, regardless of the covariance used in four tracking results. This superior performance is attributed to the tracking filter's incorporation of both measurements and CRU motion, which leads to enhanced accuracy. In contrast, snapshot positioning disregards CRU motion, and dead reckoning overlooks measurements entirely. Among the four tracking results, the optimal performance is achieved when utilizing the non-stationary \ac{MSE}, as it accurately reflects the true quality of the measurement, whereas the EECLB-LOS, the EECLB-NLOS and the stationary \ac{MSE} serve merely as estimates of the measurement covariance. {Furthermore, the least satisfactory performance is observed when utilizing the stationary \ac{MSE} as the covariance. This is due to the covariance remaining static, thereby unable to accommodate fluctuations in measurement quality across time/space. When employing the EECLB-LOS, the tracking filter exhibits better performance compared to employing EECLB-NLOS. This discrepancy arises from the fact that all paths within $\mathcal{G}_{k}$ are already resolved by the channel estimator. However, resolved NLoS paths falling within the resolution cell of the estimated LoS paths introduce errors in ECLB estimation, resulting in degraded tracking performance.}


\begin{figure}
\center
%
%
\definecolor{mycolor1}{rgb}{0.00000,0.44700,0.74100}%
\definecolor{mycolor2}{rgb}{0.85000,0.32500,0.09800}%
\definecolor{mycolor3}{rgb}{0.49400,0.18400,0.55600}%
\definecolor{mycolor4}{rgb}{0.46600,0.67400,0.18800}%
\definecolor{mycolor5}{rgb}{0.30100,0.74500,0.93300}%
\begin{tikzpicture}[scale=1\columnwidth/10cm,font=\footnotesize]
\begin{axis}[%
width=8cm,
height=4cm,
at={(0in,0in)},
scale only axis,
unbounded coords=jump,
xmin=-70,
xmax=70,
xlabel style={font=\color{white!15!black}},
xlabel={Position of the vehicle on the y-axis [m]},
ymode=log,
ymin=0.01,
ymax=10,
ylabel style={font=\color{white!15!black}},
ylabel={RMSE on ToA [m]},
yminorticks=true,
xmajorgrids,
ymajorgrids,
legend style={at={(0.01,0.99)},  anchor=north west, legend cell align=left, align=left, draw=white!15!black}
]
\addplot [color=mycolor1, dotted, line width=1.5pt]
  table[row sep=crcr]{%
-70	0.453002533527898\\
-68.6	0.450993244760322\\
-67.2	0.461108118128372\\
-65.8	0.821225429404448\\
-64.4	0.139475518361226\\
-63	0.270834288538919\\
-61.6	0.167262478892051\\
-60.2	0.148141451222222\\
-58.8	0.0697868386509315\\
-57.4	0.442135200223661\\
-56	0.152479101905199\\
-54.6	0.115177938249691\\
-53.2	0.31168835228937\\
-51.8	0.623477232033906\\
-50.4	0.204618000935494\\
-49	0.459482230390299\\
-47.6	0.316384134032908\\
-46.2	0.135352430584271\\
-44.8	0.100432485584511\\
-43.4	0.0988489254566178\\
-42	0.460222395793937\\
-40.6	0.250057543379467\\
-39.2	0.522296966955693\\
-37.8	0.111463114305243\\
-36.4	0.0333633475688092\\
-35	0.0316112080493937\\
-33.6	0.141793358564319\\
-32.2	0.195918893665165\\
-30.8	0.13163226240564\\
-29.4	0.124222237691935\\
-28	0.12861966701001\\
-26.6	0.0436036959906954\\
-25.2	0.574599832249638\\
-23.8	1.05013400400135\\
-22.4	0.148483492736711\\
-21	0.0981511593738193\\
-19.6	0.146024311878339\\
-18.2	0.0352059713242082\\
-16.8	0.0362431595884098\\
-15.4	0.0636308101442821\\
-14	0.0436421926415187\\
-12.6	0.0554089097959483\\
-11.2	0.127459348733702\\
-9.8	1.03510275528707\\
-8.4	0.875931067432703\\
-7	0.14379517287218\\
-5.6	0.0578273201485238\\
-4.2	0.260986561853528\\
-2.8	1.48711582852701\\
-1.4	0.510898619715707\\
-2.20126e-11	0.456727119581914\\
1.4	0.108147034672126\\
2.8	0.103827220597436\\
4.2	0.0263739323102885\\
5.6	0.0576276038547399\\
7	0.0776711497353823\\
8.4	0.742632294537707\\
9.8	0.236613686587448\\
11.2	0.0298723252013619\\
12.6	0.11440467232953\\
14	0.0142491390998138\\
15.4	0.0102046037052457\\
16.8	0.0140183034010197\\
18.2	0.0311389667912012\\
19.6	0.0582157057480052\\
21	0.141255586163788\\
22.4	0.222393706252203\\
23.8	3.69959913783471\\
25.2	3.61438331391818\\
26.6	0.610641870953019\\
28	0.217017596397921\\
29.4	1.61210907421205\\
30.8	0.274324653252869\\
32.2	0.0881597472065115\\
33.6	0.10548996155418\\
35	0.21751378322926\\
36.4	0.164575154545658\\
37.8	1.20778383700286\\
39.2	0.941269819637955\\
40.6	0.0740580603776432\\
42	0.121472866750146\\
43.4	0.13933666241354\\
44.8	0.774565577384606\\
46.2	0.262350112950098\\
47.6	2.01094995171691\\
49	1.04750552540505\\
50.4	0.101112391521013\\
51.8	0.686688758665621\\
53.2	0.411322005209683\\
54.6	0.269226456817685\\
56	0.631099861484569\\
57.4	0.133790444416885\\
58.8	0.310467669942991\\
60.2	0.222073927965111\\
61.6	0.972666992952173\\
63	0.513407871094391\\
64.4	0.18004971534619\\
65.8	0.381982652264083\\
67.2	0.392593111498829\\
68.6	0.311569163984693\\
70	0.272054430186559\\
};
\addlegendentry{Without gating}

\addplot [color=mycolor2, line width=1.5pt]
  table[row sep=crcr]{%
-70	0.208839655273465\\
-68.6	0.225226773941074\\
-67.2	0.282937723436035\\
-65.8	0.152561970662422\\
-64.4	0.133400616295333\\
-63	0.208114082308918\\
-61.6	0.155368984247916\\
-60.2	0.138058874301185\\
-58.8	0.0697868386509315\\
-57.4	0.292646941420679\\
-56	0.15102991550826\\
-54.6	0.115177938249691\\
-53.2	0.125010781805525\\
-51.8	nan\\
-50.4	0.0791782690833137\\
-49	0.231767555418466\\
-47.6	0.290626377527248\\
-46.2	0.133450433447824\\
-44.8	0.100432485584511\\
-43.4	0.0956944950523794\\
-42	0.136366976029577\\
-40.6	0.216476291270698\\
-39.2	0.241750292601224\\
-37.8	0.111463114305243\\
-36.4	0.0333633475688092\\
-35	0.0316112080493937\\
-33.6	0.0786878350487131\\
-32.2	0.195918893665165\\
-30.8	0.13163226240564\\
-29.4	0.124222237691935\\
-28	0.115304055878744\\
-26.6	0.0436036959906954\\
-25.2	0.286091318026081\\
-23.8	nan\\
-22.4	0.148483492736711\\
-21	0.0981511593738193\\
-19.6	0.146024311878339\\
-18.2	0.0352059713242082\\
-16.8	0.0362431595884098\\
-15.4	0.0636308101442821\\
-14	0.0436421926415187\\
-12.6	0.0554089097959483\\
-11.2	0.127459348733702\\
-9.8	0.169475175792237\\
-8.4	nan\\
-7	0.14379517287218\\
-5.6	0.0578273201485238\\
-4.2	0.23846764128898\\
-2.8	0.153241261156123\\
-1.4	0.243246220214574\\
-2.20126e-11	0.171871443381124\\
1.4	0.103451660060784\\
2.8	0.103827220597436\\
4.2	0.0263739323102885\\
5.6	0.0576276038547399\\
7	0.0776711497353823\\
8.4	0.203948227180201\\
9.8	0.169016357479215\\
11.2	0.0298723252013619\\
12.6	0.095505494287846\\
14	0.0142491390998138\\
15.4	0.0102046037052457\\
16.8	0.0140183034010197\\
18.2	0.0311389667912012\\
19.6	0.0582157057480052\\
21	0.141255586163788\\
22.4	0.18870509843663\\
23.8	nan\\
25.2	0.131300065909227\\
26.6	nan\\
28	0.181279419536591\\
29.4	0.159096334332271\\
30.8	0.21363444596708\\
32.2	0.0881597472065115\\
33.6	0.10548996155418\\
35	0.215798213397643\\
36.4	0.0813531832495229\\
37.8	0.0826978120176402\\
39.2	0.233406589277159\\
40.6	0.0740580603776432\\
42	0.121472866750146\\
43.4	0.0789571925382532\\
44.8	0.240619635432878\\
46.2	0.20408844414022\\
47.6	0.260984066773495\\
49	0.244553257251695\\
50.4	0.0879818063345733\\
51.8	0.159717488351941\\
53.2	0.332591421032405\\
54.6	0.200080982559828\\
56	0.226297085475614\\
57.4	0.0850474939250646\\
58.8	0.177485455794429\\
60.2	0.0710716636742181\\
61.6	0.319684621559526\\
63	0.344210306415967\\
64.4	0.164416224383583\\
65.8	0.297544638108666\\
67.2	0.252467644985468\\
68.6	0.276481476363394\\
70	0.208631768837074\\
};
\addlegendentry{Gating}

\end{axis}
\end{tikzpicture}%
\vspace{-3mm} 
\caption{Comparison of the \acp{RMSE} of ToA estimates for the \ac{LoS} path in both cases, with and without gating.}
\label{rmse_gating}
\vspace{-3mm} 
\end{figure}
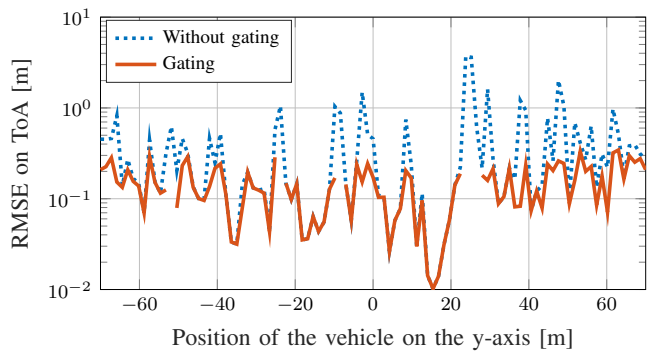

\begin{figure}
\center
\input{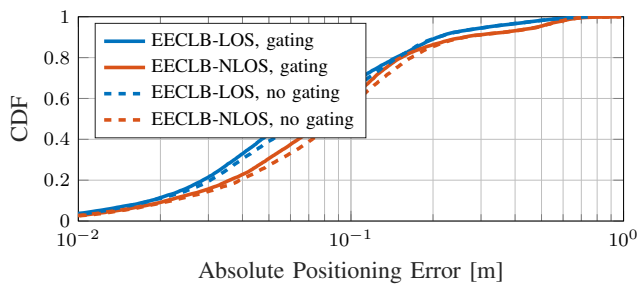}
\vspace{-3mm} 
\caption{Comparison of tracking performances when EECLB-LOS and EECLB-NLOS are used as the measurement covariance in the update step of the tracking filter in both cases, with and without gating.}
\label{cdf_gating}
\vspace{-5mm} 
\end{figure}

Next, we investigate how gating enhances the tracking performance by aiding the \ac{LoS} identification. Fig.~\ref{rmse_gating} displays \acp{RMSE} of ToA estimates for the \ac{LoS} path in both gating and non-gating cases. The lower \ac{RMSE} in the gating case is attributed to the correct identification of the \ac{LoS} path among the detected paths with gating assistance. The discontinuities observed in the solid line, e.g, around $y=24~\text{m}$ in Fig.~\ref{rmse_gating}, correspond to instances where no LoS measurements are identified (i.e., no measurements fall within the LoS gate). In such instances, the dashed line generally exhibits larger errors, because of incorrect identification of the LoS path among the detected paths. We also examine the \acp{CDF} of the absolute positioning error for the tracking results using two EECLBs, in cases with and without implementing gating, as illustrated in Fig.~\ref{cdf_gating}. Improved tracking outcomes are observed in both cases when either the EECLB-LOS or EECLB-NLOS is used, as indicated by the dashed lines positioned to the right of the solid lines in Fig.~\ref{cdf_gating}. This improvement stems from the correct identification of the LoS path among the detected paths in certain instances facilitated by gating. Conversely, without gating, the LoS path is often misidentified.

\vspace{0mm} 
\section{Conclusions}
In this paper, our focus lies on addressing the V2X sidelink tracking problem within the framework of 3GPP Release 18 in sub-6 GHz. Our objective is to track the trajectory of a moving \ac{CRU} over time, aided by a single \ac{RSU}. We develop a tracking filter based on the \ac{KF}, wherein two EECLBs are  employed as the measurement covariance in the update step of the tracking filter. Additionally, we introduce a gating method wherein the intermediate tracking results are leveraged to enhance LoS identification. To evaluate the efficacy of our approach, we conduct simulations in an urban scenario at an intersection using ray-tracing data. Our results demonstrate that the tracking filter, coupled with the proposed EECLBs, exhibits satisfactory performance, albeit with a discernible gap compared to the case employing the non-stationary \ac{MSE}. Furthermore, our results underscore the efficacy of the proposed gating method in improving overall tracking performance by enabling more accurate identification of LoS paths among the detected paths.


\vspace{0mm}
\section*{Acknowledgments}
The authors wish to thank Remcom for providing Wireless InSite\textregistered  ray-tracer.

\vspace{-4.5mm}

\balance

\bibliography{IEEEabrv_paper,Bibl_paper}

\begin{thebibliography}{10}
\providecommand{\url}[1]{#1}
\csname url@rmstyle\endcsname
\providecommand{\newblock}{\relax}
\providecommand{\bibinfo}[2]{#2}
\providecommand\BIBentrySTDinterwordspacing{\spaceskip=0pt\relax}
\providecommand\BIBentryALTinterwordstretchfactor{4}
\providecommand\BIBentryALTinterwordspacing{\spaceskip=\fontdimen2\font plus
\BIBentryALTinterwordstretchfactor\fontdimen3\font minus \fontdimen4\font\relax}
\providecommand\BIBforeignlanguage[2]{{%
\expandafter\ifx\csname l@#1\endcsname\relax
\typeout{** WARNING: IEEEtran.bst: No hyphenation pattern has been}%
\typeout{** loaded for the language `#1'. Using the pattern for}%
\typeout{** the default language instead.}%
\else
\language=\csname l@#1\endcsname
\fi
#2}}
\renewcommand\BIBentryALTinterwordstretchfactor{4}

\bibitem{wild2021joint}
T.~Wild, \emph{et~al.}, ``Joint design of communication and sensing for beyond {5G} and {6G} systems,'' \emph{IEEE Access}, vol.~9, pp. 30\,845--30\,857, 2021.

\bibitem{liu2022integrated}
F.~Liu, \emph{et~al.}, ``Integrated sensing and communications: {Towards} dual-functional wireless networks for {6G} and beyond,'' \emph{IEEE JSAC}, 2022.

\bibitem{KanRap:21}
O.~{Kanhere} \emph{et~al.}, ``Position location for futuristic cellular communications: {5G} and beyond,'' \emph{{IEEE} Communications Magazine}, vol.~59, no.~1, pp. 70--75, 2021.

\bibitem{hexax_d31}
\BIBentryALTinterwordspacing
H.~Wymeersch \emph{et~al.}, ``Localisation and sensing use cases and gap analysis,'' Hexa-X project Deliverable D3.1, v1.4, 2022. [Online]. Available: \url{https://hexa-x.eu/deliverables/}
\BIBentrySTDinterwordspacing

\bibitem{wild20236g}
T.~Wild, \emph{et~al.}, ``{6G} integrated sensing and communication: {From} vision to realization,'' in \emph{European Radar Conference (EuRAD)}, 2023, pp. 355--358.

\bibitem{tr:38845-3gpp21}
3GPP, ``Study on scenarios and requirements of in-coverage, partial coverage, and out-of-coverage {NR} positioning use cases,'' TR 38.845, Technical Report 17.0.0, 2021.

\bibitem{5g_nr_v2x_driving}
H.~Bagheri, \emph{et~al.}, ``{5G NR-V2X}: {Toward} connected and cooperative autonomous driving,'' \emph{IEEE Communications Standards Magazine}, vol.~5, no.~1, pp. 48--54, 2021.

\bibitem{garcia2021tutorial}
M.~H.~C. Garcia, \emph{et~al.}, ``A tutorial on {5G} {NR} {V2X} communications,'' \emph{IEEE Communications Surveys \& Tutorials}, vol.~23, no.~3, pp. 1972--2026, 2021.

\bibitem{lien20203gpp}
S.-Y. Lien, \emph{et~al.}, ``{3GPP} {NR} sidelink transmissions toward {5G} {V2X},'' \emph{IEEE Access}, vol.~8, pp. 35\,368--35\,382, 2020.

\bibitem{tr:38859-3gpp22}
3GPP, ``Study on expanded and improved {NR} positioning,'' TR 38.859, Technical Report 0.1.0, 2022.

\bibitem{ko2021v2x}
S.-W. Ko, \emph{et~al.}, ``{V2X}-based vehicular positioning: {Opportunities}, challenges, and future directions,'' \emph{IEEE Wireless Communications}, vol.~28, no.~2, pp. 144--151, 2021.

\bibitem{liu2021highly}
Q.~Liu, \emph{et~al.}, ``A highly accurate positioning solution for {C-V2X} systems,'' \emph{Sensors}, vol.~21, no.~4, p. 1175, 2021.

\bibitem{fouda2021dynamic}
A.~Fouda, \emph{et~al.}, ``Dynamic selective positioning for high-precision accuracy in {5G NR} {V2X} networks,'' in \emph{IEEE Vehicular Technology Conference (VTC2021-Spring)}, 2021.

\bibitem{liu2021v2x}
Q.~Liu, \emph{et~al.}, ``A {V2X}-integrated positioning methodology in ultradense networks,'' \emph{IEEE Internet of Things Journal}, vol.~8, no.~23, pp. 17\,014--17\,028, 2021.

\bibitem{ge2023analysis}
Y.~Ge, \emph{et~al.}, ``Analysis of {V2X} sidelink positioning in sub-6 {GHz},'' in \emph{IEEE International Symposium on Joint Communications \& Sensing (JC\&S)}, 2023, pp. 1--6.

\bibitem{ge2024v2x}
Y.~Ge, \emph{et~al.}, ``{V2X} sidelink positioning in {FR1}: {Scenarios}, algorithms, and performance evaluation,'' \emph{IEEE Journal on Selected Areas in Communications}, 2024.

\bibitem{koivisto2017jointd}
M.~Koivisto, \emph{et~al.}, ``Joint device positioning and clock synchronization in {5G} ultra-dense networks,'' \emph{IEEE Transactions on Wireless Communications}, vol.~16, no.~5, pp. 2866--2881, 2017.

\bibitem{kadambi2022neural}
S.~Kadambi, \emph{et~al.}, ``Neural {RF} {SLAM} for unsupervised positioning and mapping with channel state information,'' in \emph{IEEE International Conference on Communications}, 2022, pp. 3238--3244.

\bibitem{karfakis2023nr5g}
P.~T. Karfakis, \emph{et~al.}, ``{NR5G-SAM}: {A SLAM} framework for field robot applications based on {5G} new radio,'' \emph{Sensors}, vol.~23, no.~11, p. 5354, 2023.

\bibitem{talvitie2023orientation}
J.~Talvitie, \emph{et~al.}, ``Orientation and location tracking of {XR} devices: {5G} carrier phase-based methods,'' \emph{IEEE Journal of Selected Topics in Signal Processing}, 2023.

\bibitem{gustafsson2010statistical}
F.~Gustafsson, \emph{Statistical Sensor Fusion}.\hskip 1em plus 0.5em minus 0.4em\relax Studentlitteratur, 2010.

\bibitem{heath2016overview}
R.~W. Heath, \emph{et~al.}, ``An overview of signal processing techniques for millimeter wave {MIMO} systems,'' \emph{IEEE Journal of Selected Topics in Signal Processing}, vol.~10, no.~3, pp. 436--453, 2016.

\bibitem{5g_nr_rel16}
S.~Parkvall, \emph{et~al.}, ``{5G NR} release 16: {Start} of the {5G} evolution,'' \emph{IEEE Communications Standards Magazine}, vol.~4, no.~4, pp. 56--63, 2020.

\bibitem{van2004detection}
H.~L. Van~Trees, \emph{Detection, Estimation, and Dodulation Theory, Part I: Detection, Estimation, and Linear Modulation Theory}.\hskip 1em plus 0.5em minus 0.4em\relax John Wiley \& Sons, 2004.

\bibitem{tr:37885-3gpp19}
3GPP, ``Technical specification group radio access network; study on evaluation methodology of new vehicle-to-everything ({V2X}) use cases for {LTE} and {NR},'' TR 37.885, Technical Report 15.3.0, 2019.

\bibitem{WirelessInSite}
\BIBentryALTinterwordspacing
{Remcom}. {Wireless {InSite} - {3D} Wireless Prediction Software}. Accessed: Oct 9, 2022). [Online]. Available: \url{https://www.remcom.com/wireless-insite-em-propagation-software}
\BIBentrySTDinterwordspacing

\end{thebibliography}

\end{document}